\documentstyle[preprint,aps,prl]{revtex}
\begin{document}
\baselineskip=0.8cm
\title
{Quantum Monte Carlo evidence for superconductivity
in the three-band Hubbard model in two dimensions}
\author{Kazuhiko Kuroki and Hideo Aoki}
\address{Department  of  Physics, University of Tokyo, Hongo,
Tokyo  113, Japan}
\maketitle
\begin{abstract}
A possibility of the
electronic origin of the high-temperature superconductivity in cuprates
is probed with the
quantum Monte Carlo method by revisiting the three-band Hubbard model
comprising Cu$3d_{x^2-y^2}$ and O$2p_\sigma$ orbitals.
The $d_{x^2-y^2}$ pairing correlation
is found to turn into an increasing function of the
repulsion $U_d$ within the $d$ orbitals or the $d$-$p$ level off-set
$\Delta \varepsilon$,
where the correlation grows with the system size.
We have detected this in both the charge-transfer and Mott-Hubbard regimes
upon entering the strong-correlation region
($U_d$ or $\Delta \varepsilon >$ bare band width).

\end{abstract}
\ \\
\newpage
The high-temperature superconductivity
in cuprates harbors some of the most fascinating aspects of
strongly-correlated electron systems, but, despite a body of theoretical works
on possible electronic mechanisms of superconductivity,
we are still someway from a complete understanding of what happens
in the realistic parameter range.
Experimental and theoretical studies have indicated that the essence of
the cuprates lies in the two-dimensional
CuO$_2$ plane, for which it is generally recognized
that Emery's three-band Hubbard model \cite{Emery}
is the basic, starting model
that describes both
the copper $3d$ and oxygen $2p_x$ and $2p_y$ orbitals.

The model captures the essential feature of the system with
two key parameters:
$U_d$ (the on-site Coulomb repulsion between copper $d$ holes) and
$\Delta\varepsilon$ (Cu$3d$-O$2p$ level offset), where
the energies are measured in units of the
$d$-$p$ hybridization, $t_{dp}$.
The presence of both parameters makes the physics richer.
Specifically,
the inequality $\Delta\varepsilon < U_d$
is usually used to identify the insulating host material
as a charge-transfer insulator, as opposed to the Mott-Hubbard
insulator with $\Delta\varepsilon > U_d$\cite{Zaanen}.  Here
we shall extend this terminology into the doped case.
The three-band Hubbard Hamiltonian is given in standard notations as
\begin{eqnarray}
{\cal H}&=&t_{dp}\sum_{\langle i,j\rangle \sigma}(d_{i \sigma}^\dagger
p_{j \sigma}+{\rm h.c.})
+t_{pp}\sum_{\langle j,j'\rangle \sigma}(p_{j \sigma}^\dagger p_{j' \sigma}+
{\rm h.c.})\nonumber\\
&+&\Delta\varepsilon\sum_{j \sigma}n_{j \sigma}^p+
U_d\sum_i n_{i\uparrow}^d n_{i\downarrow}^d
\end{eqnarray}
where
$d^\dagger$ creates a Cu3$d_{x^2-y^2}$ hole and $p^\dagger$ an
O2$p_\sigma$ hole,
$t_{dp} (t_{pp})$ is the nearest-neighbor
$d$-$p$ ($p$-$p$) transfer. Here the repulsion within
the $p$ orbitals and the repulsion between $d$ and $p$ orbitals
have been neglected for simplicity.

Great efforts have been made to search for
superconductivity in this model\cite{Imada,Scalettar,Dopf,Assaad},
but indications of the off-diagonal long-range order have not
been detected so far.
There is also a variational Monte Carlo study\cite{Coppersm}, but the
justification of the variational wave functions remains somewhat open.

Subsequently reductions to
effective Hamiltonians as certain limits of the original
three-band model have been attempted.  In the limit of large level offset
($\Delta\varepsilon\gg U_d, t_{dp}$), the system is equivalent to the
{\it single-band} Hubbard model with the on-site interaction $U_d$ and the
effective nearest-neighbor hopping $t_{\rm eff}=t_{dp}^2/\Delta\varepsilon$.
If we further put $U\gg t_{\rm eff}$,
the system reduces to the $t$-$J$ model
with excluded double occupancies
and $J=4t_{\rm eff}^2/U$.
Thus the $t$-$J$ model is a natural limit of the
three-band model in the Mott-Hubbard regime.

However, the real cuprates lie in the charge-transfer
regime.
Zhang and Rice \cite{Zhang} have proposed
that even in this case, the low-lying states of the
three-band model may be essentially
represented by the $t$-$J$ model, at least in the limit of
$U_d \gg \Delta\varepsilon \gg t_{dp}$, and provided that
the spin-triplet $d$-$p$
molecular orbitals may be neglected.
The $t$-$J$ model thus picks up spin-singlet $d$-$p$
molecular orbitals (or bonding Kondo states), which are envisaged to
experience
a superexchange interaction $J$ while moving around with
an effective transfer $t$ with double occupancies avoided.
The superexchange provides a natural source of an effective attraction
among the molecular orbitals, and
extensive theoretical works
have indeed indicated that the $t$-$J$ model
superconducts for a certain range of $J/t$.
In one
dimension (1D) this is shown clearly
from a phase diagram having a finite pairing-dominated region around
$J\sim 2t$\cite{Ogata}.
In 2D, exact diagonalization
results\cite{Dagotto} indicate that
the $d_{x^2-y^2}$-wave paring correlation function is long-tailed
for sufficiently large $J\sim t$, which is also supported from
variational Monte Carlo studies\cite{Gros,Yokoyama,Giamarchi}.

Now, even if the $t$-$J$ model can be superconductive,
the following fundamental questions do remain for the original
three-band Hubbard model:

\noindent (i) Does the
perturbative picture that maps the three-band model into $t$-$J$ model
in the limit of $t_{dp}/\Delta\varepsilon,
t_{dp}/U_d \rightarrow 0$
remain valid for finite, realistic values of parameters ?
In real materials $\Delta\varepsilon\sim 2.5t_{dp}$ \cite{Hybertsen}
is only moderate,
where the validity of the perturbation is not at all clear.

\noindent
(ii) Even if the perturbation is to remain valid
through e.g. renormalizations,
whether the resultant $J/t$ can become large enough
to guarantee a high $T_C$ is a nontrivial question.

\noindent
(iii) Would there be a qualitative difference between the
Mott-Hubbard and charge-transfer regimes
concerning the appearance of superconductivity via e.g. different effective
$J/t$ mentioned in (ii) ?

All these points evoke another basic question, i.e.,
does the {\em single-band} Hubbard model, which shares the $t$-$J$ model
as an effective Hamiltonian in the strong-correlation limit, have a
superconducting phase ?
In 1D, the conformal field theory indicates
that no matter how $U/t$ is increased,  the
superconducting correlation fails to become dominant,\cite{Kawakami,Schulz}
indicating a behavior distinct from the situation when
we let $J\sim t$ in the $t$-$J$ model.
In the 2D Hubbard model, quantum Monte Carlo calculations
up to $U=4t$ still show no sign of the
off-diagonal long-range order\cite{Furukawa}.
To reconcile this, we have to consider a possibility that
either the effective $J/t$ is small,
or $U=4t$ is already
outside the perturbative region.
If the single-band Hubbard model remains normal for
the whole range of parameters, while the three-band
Hubbard model with finite, realistic values of parameters
does superconduct, the
Mott-Hubbard and charge-transfer regimes may possibly
belong to different universality classes.

These problems have remained a long-standing puzzle,
which is exactly our motivation to revisit
the three-band Hubbard model, where we cover
a hitherto unfathomed range of parameters.
If the answer is positive, we will have a stronger ground to consider the
superconductivity in cuprates
to be of electronic origin.

We employ the quantum Monte Carlo (QMC) method, where
our motivation described above calls for special emphasis upon the following.

\noindent (i) We consider the range of
$\Delta\varepsilon$ and $U_d$ extending to the bare width, $W$, of the
most relevant (Cu$3d$-O$2p_\sigma$ anti-bonding) band.
We define the case where
both $\Delta\varepsilon$ and $U_d$ are comparable with $W$
to be the strong-correlation regime in the
following sense.
The relevant energy to be compared with $W$ should be the effective repulsion
within the $d$-$p$ Wannier orbital, which should be greater than
${\rm Min} \{\Delta\varepsilon, U_d\}$,
the minimum cost of energy for two holes occupying the same
Wannier orbital.  This is in fact illustrated in the low-lying spectra of
finite systems,
where the levels of the three-band model with
$\Delta\varepsilon=3.6$eV and $U_d=10.5$eV are best-fit to
those of the single-band
Hubbard model with $U\sim 5$eV\cite{Hybertsen}.

\noindent (ii) The carrier doping is kept
close to the experimentally known
optimum value ($\delta \sim 0.15$) for the superconductivity.

\noindent (iii) Since a reliable detection of the
pairing correlation is required, we adopt the ground-state
(or projector) QMC formalism with the projection imaginary
time of at least $12t_{dp}$ to ensure convergence.

\noindent (iv) The sample-size dependence is studied for lattice sizes
up to $8\times 8$ unit cells (192 atoms), which is combined with a
real-space analysis to probe the range of the pairing correlation.

To our knowledge, previous calculations do not
satisfy all of these conditions simultaneously.
For the largest $\Delta\varepsilon$ and $U_d$ considered here,
the CPU time required was typically 50 hours on HITAC S-3800 supercomputer.

As for the symmetry of the pairing, we have considered
$d_{x^2-y^2}$-wave $(f_d=\cos q_x-\cos q_y)$
and extended $s$-wave $(f_s=\cos q_x+\cos q_y)$
pairing, for which we have calculated
the $k=0$ Fourier component of the real space correlation function,
$S_\alpha=\frac{1}{2N}\langle \Delta_\alpha^\dagger\Delta_\alpha+
\Delta_\alpha\Delta_\alpha^\dagger\rangle$, with
$
\Delta_\alpha=\sum_{\bf q}f_\alpha({\bf q})
(d_{{\bf q}\uparrow}d_{{\bf -q}\downarrow}
+p_{{\bf q}\uparrow}^x p_{{\bf -q}\downarrow}^x
+p_{{\bf q}\uparrow}^y p_{{\bf -q}\downarrow}^y).
$

We first focus on the hole doping.
We go from 18 holes for $4\times 4$ unit cells
(doping $\delta = 0.125$),
42 for $6\times 6$
($\delta = 0.166$), to
74 for $8\times 8$
($\delta = 0.156$).
These fillings are chosen so as to satisfy (i) the proximity to
$\delta\sim 0.15$, and
(ii) the closed-shell condition (with a non-degenerate one-electron
ground state) to ensure the stability of the QMC calculation.
We have set $t_{pp}=-0.4t_{dp}$\cite{Hybertsen}.

In Fig.\ref{fig1} the dependence
of $S_d$ on $\Delta\varepsilon$ (a) or $U_d$ (b) is shown.
For small $\Delta\varepsilon$ and/or $U_d$, $S_d$
decreases with $\Delta\varepsilon$ and $U_d$.
An increase in $\Delta\varepsilon$ or $U_d$ implies an increased
ratio (electron-electron repulsion)/(band width),
and the result in the weakly correlated regime
shows that this works unfavorably for superconductivity
as naively expected.
However, $S_d$ dramatically begins to increase with these parameters
for larger values of $\Delta\varepsilon$ and/or $U_d$.
The crossover to this behavior occurs in the
`strong-correlation' regime
where both $\Delta\varepsilon$ and $U_d$ exceed
the band width $W$ of the anti-bonding $d$-$p$ band ($W\sim 2.33t_{dp}$ for
$\Delta\varepsilon=2.7t_{dp}$ and $t_{pp}=-0.4t_{dp}$).

Right above the strong-correlation regime
the pairing correlation starts to {\em grow} with the
system size, which can be interpreted
as a tendency toward the formation of off-diagonal long-range order.
This is in sharp contrast with
the weak-correlation
regime, where $S_d$ has a small, inverse size dependence.

To check that we are really looking at the long-range part of the
pairing correlation, we have looked into their
behavior in a real space.
If we decompose $S_\alpha$ into a sum over
the real space distance $\Delta {\bf r}$,
$S_\alpha=\sum_{\Delta {\bf r}}s_\alpha(\Delta {\bf r})$
with $s_\alpha(\Delta {\bf r})$ being the
correlation function in real space.
In Fig.\ref{fig2} we represent $s_d(\Delta {\bf r})$  by
$S_d(R)$ defined by
restricting the sum in the above formula to $|\Delta x|, |\Delta y|\leq R$
(in the periodic boundary condition)
, where ${\Delta\bf r}=(\Delta x,\Delta y)$.
We can see that $S_d(R)$ monotonically increases as
we include more and more distant correlations,
which implies that the growth of the $k=0$ component, $S_d$,
is indeed caused by the extension of the pairing correlation
beyond the system size.

An indication that this kind of caution is indeed necessary is shown in
the inset of Fig.\ref{fig2}.  Namely, although the extended
$s$-wave pairing correlation, $S_s$, also increases with the system size,
its real-space behavior, $S_s(R)$,
is almost a constant,
indicating that the size dependence only signifies a
short-range correlation.

Now, despite the absence of electron-hole symmetry in the
three-band Hubbard model,
the electron-doped materials such as Nd$_{1-x}$Ce$_{x}$CuO$_4$
have been shown to superconduct as well.
Quite apart from this,
the electron-doped case is of another theoretical interest in the
following sense. In the limit of
$\Delta\varepsilon\rightarrow\infty$, $U_d\rightarrow\infty$
the electrons are doped to (i.e., holes are taken out from)
the Cu$3d$ orbital, with only small amount of carriers left in O$2p$.
This should make the system closer to
the {\it single-band} Hubbard
model than in the hole-doped case.

In doping electrons
(or in taking out holes in the hole picture)
the best choice of the band filling satisfying the above conditions are 58
holes $/8\times8$ (with the doping level $\delta=0.1$) and 26 holes
$/6\times 6$ ($\delta=0.26$).
Thus, the band fillings are unfortunately not so close for the two sizes,
which makes the analysis less conclusive.

Nevertheless, $S_d$
does again become an increasing function
of $\Delta \varepsilon$ and $U_d$ for $\Delta \varepsilon\sim 2.5t_{dp}$
and $U_d\sim 3.0t_{dp}$, where $S_d$ grows
as the size becomes $8\times 8$ from $6 \times 6$,
indicating a tendency toward $d_{x^2-y^2}$ electron-pairing in
the strongly-correlated regime.

Encouraged by the electron-doped result,
we move on to our final motivation.
Namely we go back to the
hole-doped case to investigate
the Mott-Hubbard regime ($U_d<\Delta\varepsilon$)
with {\em large} $\Delta\varepsilon$,
which leaves few O$2p$ holes to give another natural way to approach
the single-band Hubbard model as mentioned earlier.

In Fig.\ref{fig3}, we show the dependence of $S_d$ on
$\Delta\varepsilon$ with a fixed $U_d=1.8t_{dp}$ (a)
or on $U_d$ with a fixed $\Delta\varepsilon=3.6t_{dp}$ (b)
with the same system sizes and band fillings
as in Fig.\ref{fig1}.
Strikingly enough, the system size dependence does appear as well
for larger $\Delta\varepsilon$ and $U$ just like in Fig.\ref{fig1}.

If we now combine this result with that in the
hole-doped charge-transfer regime,
the following picture emerges.
Suppose we compare the relevant energy in the Mott-Hubbard regime,
Min$\{\Delta\varepsilon, U_d\}=U_d$, with
the width of the anti-bonding band, which is
$W=1.87t_{dp}$ for $\Delta\varepsilon=3.6t_{dp}$ (and
$t_{pp}=-0.3t_{dp}$ which we have assumed here).
The region at which the pairing correlation emerges is
precisely $U_d\sim W$,
which is a counterpart to
$\Delta \varepsilon \sim W$ in the charge-transfer regime.
Hence, {\it no matter which regime in
the three-band model}, a tendency
toward $d_{x^2-y^2}$-wave pairing superconductivity
emerges when the relevant energy ($U_d$ or $\Delta \varepsilon$)
exceeds $W$, i.e., when
our definition of the strong-correlation criterion is met.
In principle we can thus envisage a superconductivity phase diagram against
$\Delta\varepsilon$ and $U_d$, something like
$U\Delta \varepsilon > 8t^2_{dp}$ if we use the perturbative
expression for $W$.

If we now recall our reasoning,
the result summarized above amounts that
either (i) the single-band Hubbard model
with $U/t$ as large as the bare band width should concomitantly exhibit
superconductivity,
or (ii) we are looking at a regime where
the finiteness of $\Delta
\varepsilon$ makes the universality class of the
three-band model distinct from that of the single-band Hubbard model
through e.g. different ranges of the effective $J/t$.
The former possibility that the three-band model already
resembles the one-band Hubbard model
when $\Delta\varepsilon$ is increased up to $3.6t_{dp}$
does not contradict with the
previous one-band QMC results,
where the largest $U$ so far studied is only half the band
width, $4t$.
$U=\frac{1}{2}W$ is mimicked by the three-band
model with $U_d \sim t_{pd}$ for $\Delta\varepsilon=3.6t_{dp}$,
for which the sign of pairing is certainly
absent in the present result as well.
We believe this problem deserves further investigations.

Finally we comment on the possible
relevance of our result in the charge-transfer regime to the high $T_C$
materials.
The value of $\Delta\varepsilon
\sim 2.5t_{dp}$ where $S_d$ grows with system size is
remarkably close to the value
$(\Delta\varepsilon=2.7t_{dp})$ obtained from a first-principles
calculation for La$_2$CuO$_4$\cite{Hybertsen}.
As for the value of $U_d$, the maximum value tractable with the QMC
$(U_d/t_{dp}= 3\sim3.5)$ happens to be smaller
than realistic values $(U_d/t_{dp} =6\sim8)$,
but even for these small values the tendency for superconductivity
already emerges.
We expect that the tendency can become stronger
for larger values of $U_d$.

In summary, we have detected an indication of superconductivity in
the three-band Hubbard model without reducing it into some effective
model.
Strikingly, this indication emerges in both of the charge-transfer and
Mott-Hubbard regimes, and
in both of the hole-doped and electron-doped cases
as long as we are sitting in the strong-correlation
regime.  Since all of these regimes and cases share the
$t$-$J$ model as  some limiting  cases,
this might suggest
a scenario in which the superconductivity
as conceived in the $t$-$J$ limit
extends well into the realistic parameter regime.

Numerical calculations were done on HITAC S3800/280 at the Computer Center
of the
University of Tokyo, and FACOM VPP 500/40 at the Supercomputer Center,
Institute for Solid State Physics, University of Tokyo. This work was
supported by `Project for Parallel Processing and Super Computing'
at Computer Centre, University of Tokyo, arranged by Prof. Y. Kanada,
and also by Grant No. 07237209
from the Ministry of Education, Science, and Culture, Japan.

\begin{figure}
\caption{The $d_{x^2-y^2}$-wave pairing correlation, $S_d$,
is plotted (a) against $\Delta\varepsilon$
for a fixed $U_d=3.2$,
and (b) against $U_d$ for a fixed $\Delta\varepsilon=2.7$.
We assume the hopping integrals $t_{dp}=1$, $t_{pp}=-0.4$.
Number of holes and the sizes of the system are
18 holes$/4\times 4$ unit cells ($\triangle$), 42 $/6\times 6$ ($\bigcirc$),
and 74 $/8\times 8$ ($\Box$).
For $8\times 8$ a wider range is displayed in the inset of (a) to
show the change in sign of the gradient.
The dashed lines are guide for the eye.
}
\label{fig1}
\end{figure}

\begin{figure}
\caption{The $d_{x^2-y^2}$-wave pairing correlation, $S_d(R)$,
and the extended $s$-wave pairing correlation, $S_s(R)$ (inset)
are plotted as a
function of the range, $R$, in real space.
}
\label{fig2}
\end{figure}

\begin{figure}
\caption{Similar plot for $S_d$ as in Fig.1 in the Mott-Hubbard regime,
$U<\Delta \varepsilon$.
$S_d$ is plotted (a) against $\Delta\varepsilon$
for a fixed $U_d=1.8$,
and (b) against $U_d$ for a fixed $\Delta\varepsilon=3.6$ for
$t_{dp}=1$, $t_{pp}=-0.3$.
}
\label{fig3}
\end{figure}
\end{document}